\documentclass[12pt]{article}
\usepackage[dvips]{graphicx}
\begin{document}

\begin{center}
{\Large\bf
Particle Acceleration and Cosmic-Ray Origin in the Galaxy\\
}
\end{center}

\baselineskip=18pt
\medskip

\centerline{\Large Toru Tanimori} 
\medskip

\begin{center}
\noindent {\large \it 
Dept. of Physics, Graduate School of Science, Kyoto University,
Kitashirakawa-Oiwakechou Sakyo-ku, Kyoto, 606-8502,  Japan}

\noindent {\it 
tanimori@cr.scphys.kyoto-u.ac.jp                
}
\end{center}

\begin{abstract}
In 1990's Very High Energy Gamma-ray Astrophysics has 
dramatically advanced due to the Imaging Air \v Cerenkov   
Telescopes(IACTs).
After the first  detection of TeV gamma-ray
emission from the Crab nebula in 1989, 
several type of TeV gamma-ray sources, Active Galactic
Nuclei(AGN), young pulsar, and  SuperNova Remnant(SNR),
have been detected.
In those discoveries, recent detections of both synchrotron X-rays and 
TeV gamma-ray emissions from several SNRs are very 
significant.
SNR has been widely believed to be an unique candidate 
of galactic cosmic-ray  origin 
since the beginning  of cosmic-ray physics,
whereas little observational evidences have been reported so far.
Those are expected to be a clue of not only the galactic 
cosmic-ray origin but also the understanding of 
the particle acceleration due to a diffusive shock.

Here I present  the recent  results 
obtained by our group, CANGAROO, about the evidences of
electron and proton acceleration in SNRs.

\end{abstract}


\baselineskip=15pt plus 1pt
\section{Introduction}

\begin{figure}
\begin{center}
\includegraphics[height=6cm]{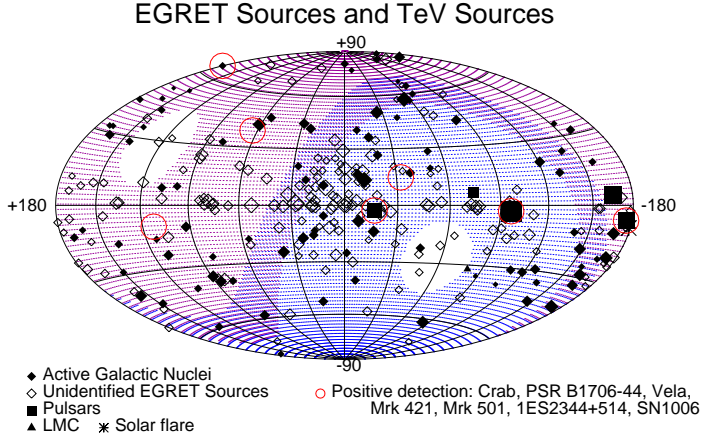}
\caption{
GeV gamma-ray source catalog detected by EGRET in
galactic coordinates, where six TeV gamma-ray sources are also drawn
by a circle.}
\label{GeVsource}
\end{center}
\end{figure}

As is well known, X-ray astronomy is an established 
field in astronomy, where
several  hundred thousands of X-rays sources have been found so far.
On the other hands, the growth of gamma-ray astronomy  had been 
relatively slow until the launch of the advanced 
gamma-ray satellite, Compton Gamma-Ray
Observatory(CGRO), in 1991\cite{ref:1}.
EGRET, a detector observing GeV gamma rays on-board CGRO,
found more than 200 gamma-ray sources, 
whereas only 20 sources had been known before then.
Emissions of high energy gamma rays were observed not only from 
galactic sources such as  pulsar/nebulae and 
supernova remnants but also the more than hundred extragalactic sources.
Figure \ref{GeVsource} 
shows the GeV gamma-ray source catalog detected by EGRET in
galactic coordinates.
You see that hundreds of AGN were located out of the galactic plane,
and moreover along the  galactic plane there concentrated lots of 
unidentified sources.

Also several TeV gamma-ray sources have been recently
established: Crab pulsar/nebula\cite{ref:2}, 
Mrk421\cite{ref:3}, Mrk501\cite{ref:4}, PSR1706-44\cite{ref:5}.
Celestial objects emitting very high energy gamma rays of energies greater
than TeV had been expected as a natural consequence
of the existence of cosmic rays, and 
they have been searched  for since the middle of 20th century
using the ground based
detectors such as scintillation-counter arrays and 
air \v Cerenkov   telescopes.
Since huge background of hadron showers overwhelms 
the tiny signal of celestial gamma rays,
no persistent TeV gamma-ray source had been found until the 
discovery of TeV gamma-ray emission from the Crab
by the Whipple group in 1989\cite{ref:2}.
The Whipple group developed imaging \v Cerenkov technique\cite{ref:6}, 
and hence the rejection power of hadron showers was greatly improved.
In 1990's several type of TeV gamma-ray sources have been 
detected with high statistics by adopting this imaging technique.
In particular,
successive discoveries of AGN emitting TeV gamma rays
were astonishing\cite{ref:3,ref:4}.

Our group, CANGAROO, the collaboration of Japanese and Australian 
institutes, 
has observed TeV gamma-ray sources in the southern hemisphere
since 1992 in South Australia\cite{ref:7}.
The southern hemisphere provides us a good chance to observe 
lots of   galactic objects such as pulsar/nebulae, supernova remnants (SNRs),
black holes, the galactic center, and so on.
In fact we have found several  galactic TeV gamma ray sources
as listed in the review article of  Weekes\cite{ref:8}. 
In particular recent noteworthy discoveries  are
the several reports on the detection
of TeV gamma-ray emissions from shell-type SNRs
in both southern and northern skies\cite{ref:9,ref:10,ref:11},
to which people have been eagerly looking forward ever since the beginning
of cosmic-ray physics.

\section{Cosmic-ray origin and shock acceleration}

High energy particles are
known to fill all over the galaxy and maybe over the surrounding halo,
and play non-negligible roles on almost all phenomena in the
universe.
Moreover, high energy particles coming to the earth, ``cosmic rays'',
are  surely affected on the circumstance of the earth.
However, nobody  know how and where such high energy particles are 
generated in the universe.
These questions are  unresolved yet, and always
important issues in astrophysics 
in spite of the long history of the study of cosmic rays.

Radio and X-ray observations
have revealed  lots of high energy phenomena in the universe,
where huge amount of energy is consumed to accelerate particles 
up to more than GeV energies,
since high energy electrons  in the  magnetic field 
emit synchrotron radiation of radio to X-ray.
Also high energy ions (mainly proton) must surely exist
in the galaxy since almost cosmic rays are
ions.

\begin{figure}
\begin{center}
\includegraphics[height=5cm]{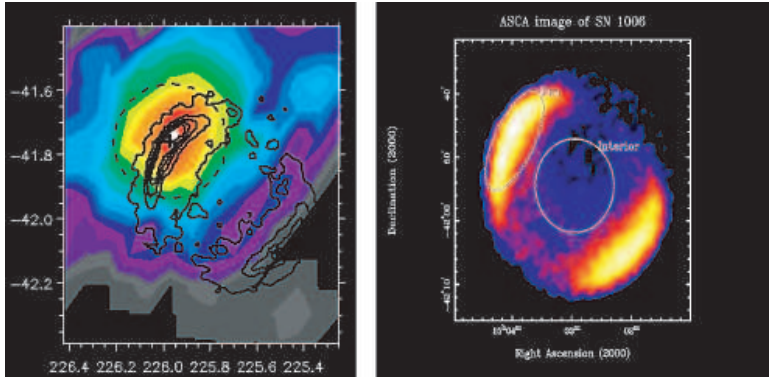}
\caption{
Map of TeV gamma-ray (right) and ASCA X-ray (left) emissions of SN1006.
}
\label{SN1006}
\end{center}
\end{figure}

For a long time,  supernova remnants (SNRs) have been believed 
to be a favored site for accelerating cosmic
rays up to 10$^{15}$eV,
because only they can satisfy the required energy input
rate to the galaxy among several galactic objects\cite{drury}.
In addition, a shock acceleration theory  was established around 1980s,
in which particles are accelerated 
with the collisions between particles and
plasma gas moving at a supersonic velocity in the space\cite{ref:13}. 
SNRs are a just extended and heated gas system 
accompanied by  very strong shocks.
Shocks are very common phenomena in the universe, and 
hence the shock acceleration  has been widely applied for high energy 
phenomena in the universe.
Thus shock-acceleration mechanism has been a standard theory
for particle acceleration in astrophysics.
Although this theory looks  very simple and reliable,
an observational evidence is still very sparse.

In order to investigate the shock-acceleration mechanism,  
SNR is an unique and ideal
laboratory because it is quite simple and 
a well-understood astronomical object.
The evolution of SNR can be  fairly explained with   
several observables such as  
explosion energy of the SNR, total mass of the ejecta, density of 
the inter stellar medium (ISM) around the SNR 
and the age  after the explosion\cite{ref:15}.  
In addition, the resoluble size of a SNR enables us to directly
observe the geometrical structure of the shock front accelerating particles, 
which provides lots
of significant physical parameters quantitatively 
(absolute value of the magnetic field, index of power law,
maximum energy of particle acceleration, and so on). 

\subsection{SN1006:first evidence of electron acceleration up to $\ge$ TeV}  

The first evidence for the very high-energy particle acceleration 
in a SNR was presented by the observation of the strong synchrotron
emission of SN1006 by the Japanese X-ray satellite ASCA in 1995\cite{ref:16}.
By assuming the magnetic field of several $\mu$
gauss, observed synchrotron X-ray emission strongly supported
the existence of high energy electrons of tens or hundreds of TeV.
Those high energy electrons must emit not only synchrotron radiation
but also high-energy gamma rays due to Inverse Compton (IC)
process by the hard collision with  2.7K Cosmic Microwave Background (CMB).
Scattered photons acquire  the energy of about one tenth of
primary electrons, and  hence their energies reach near 10 TeV in SN1006.
They  could be detected by the CANGAROO telescope in the TeV region
\cite{Pohl, Mas, Jager, Yanagita}.
In 1996 and 1997, CANGAROO succeeded in detecting the
TeV gamma-ray emission from the
north rim of SN1006\cite{ref:9} as shown in Fig.\ref{SN1006}.

\begin{figure*}
\begin{minipage}[t]{8cm}
\resizebox{8cm}{!}{\includegraphics{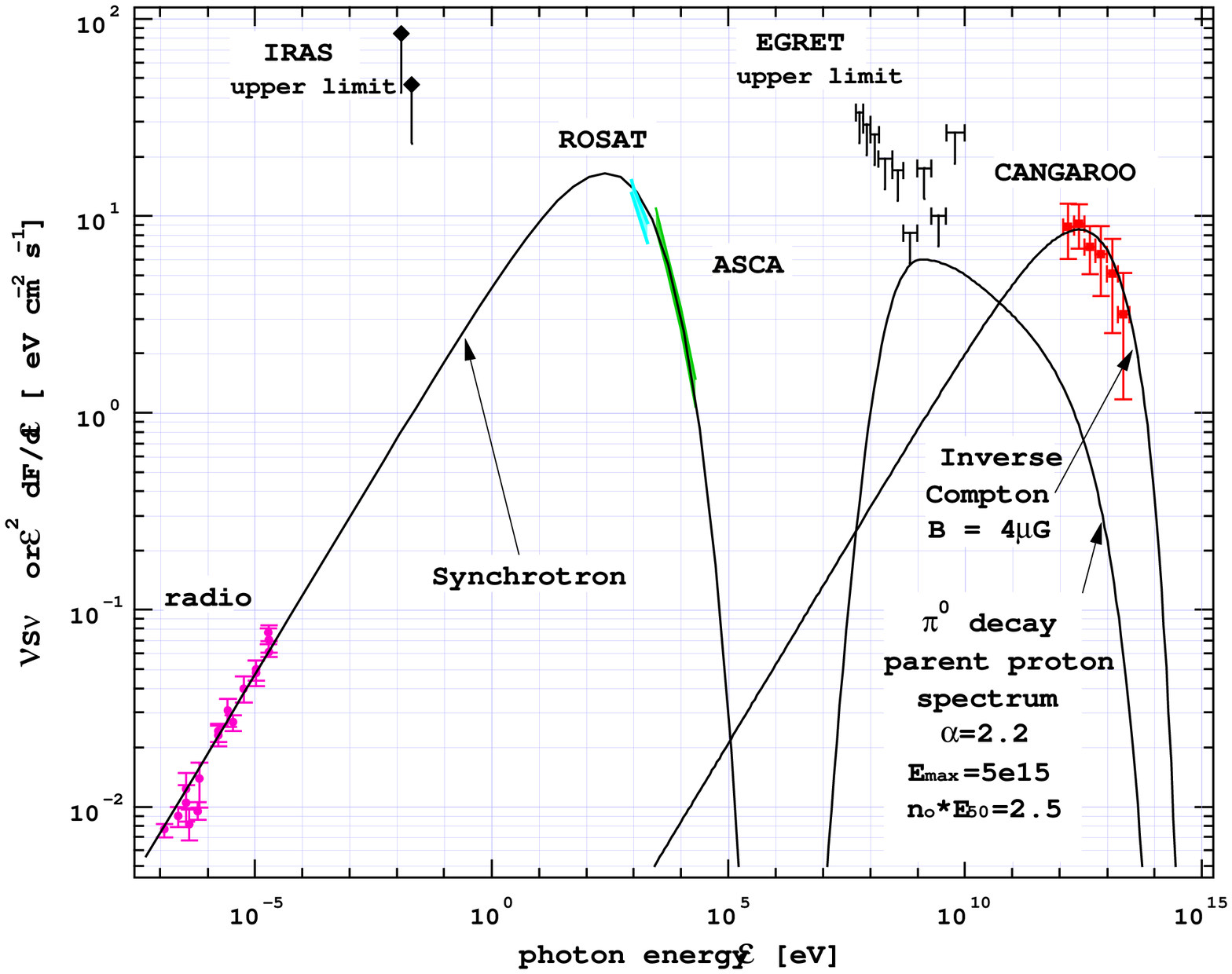}}
\caption
{{\bf a:} Energy spectrum at the north rim
of SN1006 from radio to TeV, and also fitting results based
on the shock model. 
}
\end{minipage}\hfill
\hspace{4mm}
\begin{minipage}[t]{8cm}
\resizebox{8cm}{!}{\includegraphics{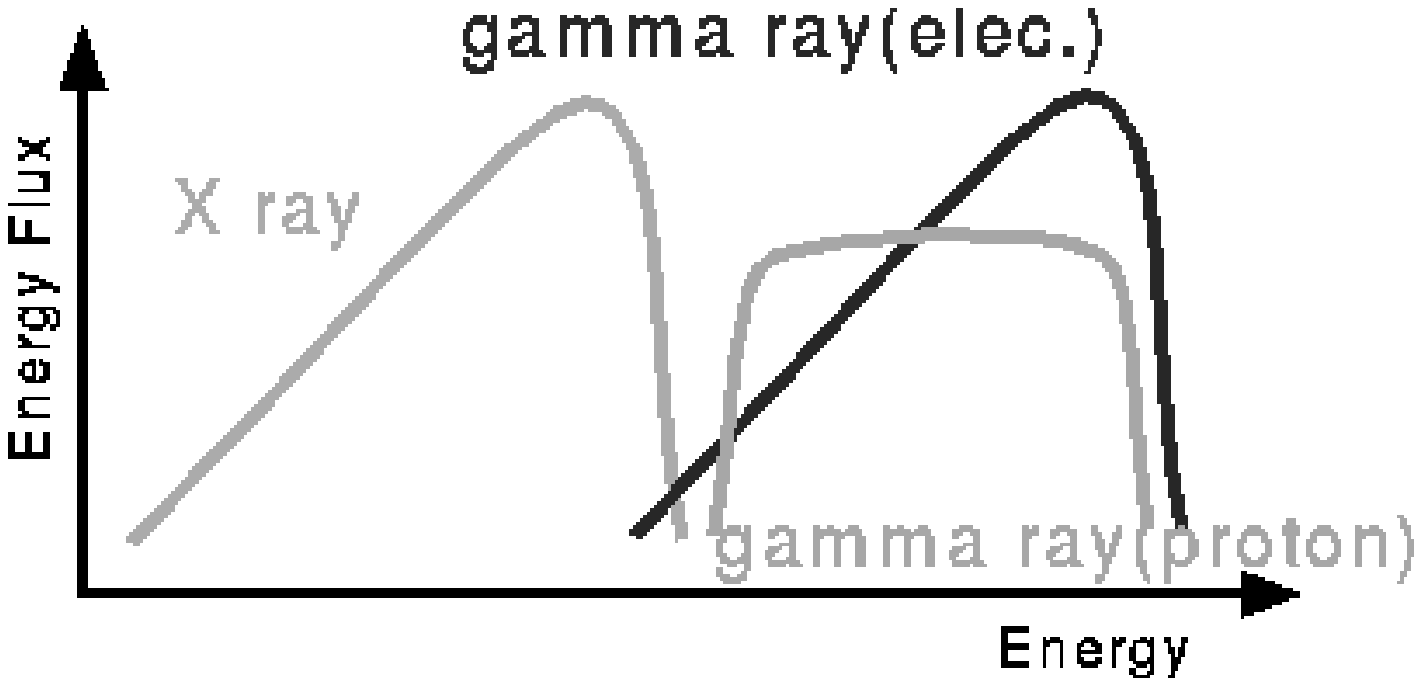}}
\caption
{{\bf b:} Expected energy spectra from both IC process and $\pi ^{\circ}$ decay
with synchrotron spectrum. 
}
\end{minipage}
\label{1006spectra}
\end{figure*}

There exists a simple but useful formula connecting  among 
relativistic electrons, high energy photons scattered by
IC process, synchrotron photons and soft seed photons. 
The emission powers of synchrotron radiation 
and IC scattering, $ P_{sync} $ and
$ P_{IC} $ are respectively expressed as follows,
\begin{equation}             
 P_{sync} = \frac{4}{3} \sigma_T c \gamma^{2} \beta^{2} U_B,
\hspace{2mm}
P_{IC} = \frac{4}{3} \sigma_T c \gamma^{2} \beta^{2} U_{soft}, 
\end{equation}
where $ \sigma_T $ is the Thomson cross section, 
$ \sigma_T = 6.7 \times 10^{-25} cm^{2} $, and
$ c\beta $ and $\gamma$ are  the velocity and Lorentz factor 
of the electron. 
$ U_B $ and $ U_{soft} $ are energy densities 
of the magnetic field and the soft seed photons, respectively. 
Since the energy density of soft seed photons (CMB) is well-known, 
an observed  gamma-ray flux  provides 
a good estimation of the magnetic field strength at the acceleration 
site in SN1006.
Figure \ref{1006spectra}a shows the wide band energy spectrum at the north rim
of SN1006 from radio to TeV, and also the fitting result based on the 
IC model\cite{ref:17}.
All data are fitted very well, and several significant 
 parameters, magnetic field({\it B}), power index(a), 
maximum energy ($ E_{max}$) 
were determined independently:  $B  = 4.3\mu $ gauss, 
a = 2.2, and $E_{max}$ = 60 TeV.

Here we assumed that the detected TeV gamma-ray emission is
mainly due to IC process with very high energy electrons
considering  the tenuous shell of ($\le \sim$0.4 cm$^{-3}$)\cite{Will}.
In fact the expected spectrum from $\pi ^{\circ}$ decay generated by
high energy proton conflicts with the upper limit in the GeV 
(Fig.\ref{1006spectra}a).

In order to verify the origin of cosmic rays,
we have to obtain a clear evidence of proton acceleration.
The identification of the parent particles of TeV gamma-rays (electron
or proton) will be possible by observing the wide spectrum from sub 
to multi
TeV region as shown in Fig.\ref{1006spectra}b.
Gamma-ray spectrum flatter than $E^{-2.0}$ in this region is surely due to
IC process, while that due to $\pi^{\circ}$ 
decay generated by
collision between intersteller matter (ISM) and high energy proton is expected to be steeper
than $E^{--2.0}$.

\section{RX J1713.7-3946: first evidence of proton acceleration up to the TeV region}     

RXJ1713.7-3946 was observed as the strongest synchrotron X-ray 
emitter among SNRs by ASCA in 1997\cite{ref:35}, and 
subsequently  TeV emission was detected from the maximum
X-ray emission point\cite{ref:10}.
Although this SNR emits an intense synchrotron X-ray similar to SN1006,
those two SNRs looks different.
Its morphology is obviously more complex than that of SN1006,
of which north parts might interact with the molecular cloud
observed by  the radio telescope\cite{ref:36}.
Therefore this TeV emission might be  ascribed to the
$\pi^{\circ}$ decay generated by the collision of accelerated
protons with the molecular cloud.

To clarify the nature of accelerated particles,
we have observed this point again in 2000 and 2001 using the new CANGAROO
10m telescope\cite{ref:ta}, 
and the result has been recently published\cite{ref:na}.
Here summary is presented.

The differential fluxes of TeV gamma-rays  from RX J1713.7-3946 are plotted in 
Fig.{\ref{1713-dif}a  with previous data\cite{ref:10}, and  
the best fit is

\begin{equation}
dF/dE = (1.63\pm 0.15\pm 0.32)\times 10^{-11}
(E/1TeV)^{-2.84 \pm 0.5 \pm 0.20}
  \,\,{\rm TeV}^{-1}{\rm cm}^{-2}{\rm s}^{-1},
\end{equation}

\noindent where the first errors are statistical 
and the second are systematic. 
You note that RX J1713.7-3946 is one of the
brightest galactic TeV gamma-ray sources discovered so far, and the power-law 
spectrum increases monotonically in a single power law as energy decreases. 
This feature is in contrast to the 
Tev gamma-ray spectrum of SN 1006, which flattens below 1 TeV\cite{n14},
and is well consistent with 
synchrotron/Inverse Compton (IC) models\cite{Pohl, Mas, Jager, Yanagita, n18}. 
While both SNRs emit intense X-rays via the synchrotron process, 
different TeV spectra suggest that different emission 
mechanisms may act respectively.

\begin{figure*}
\begin{minipage}[t]{7cm}
\resizebox{7cm}{!}{\includegraphics{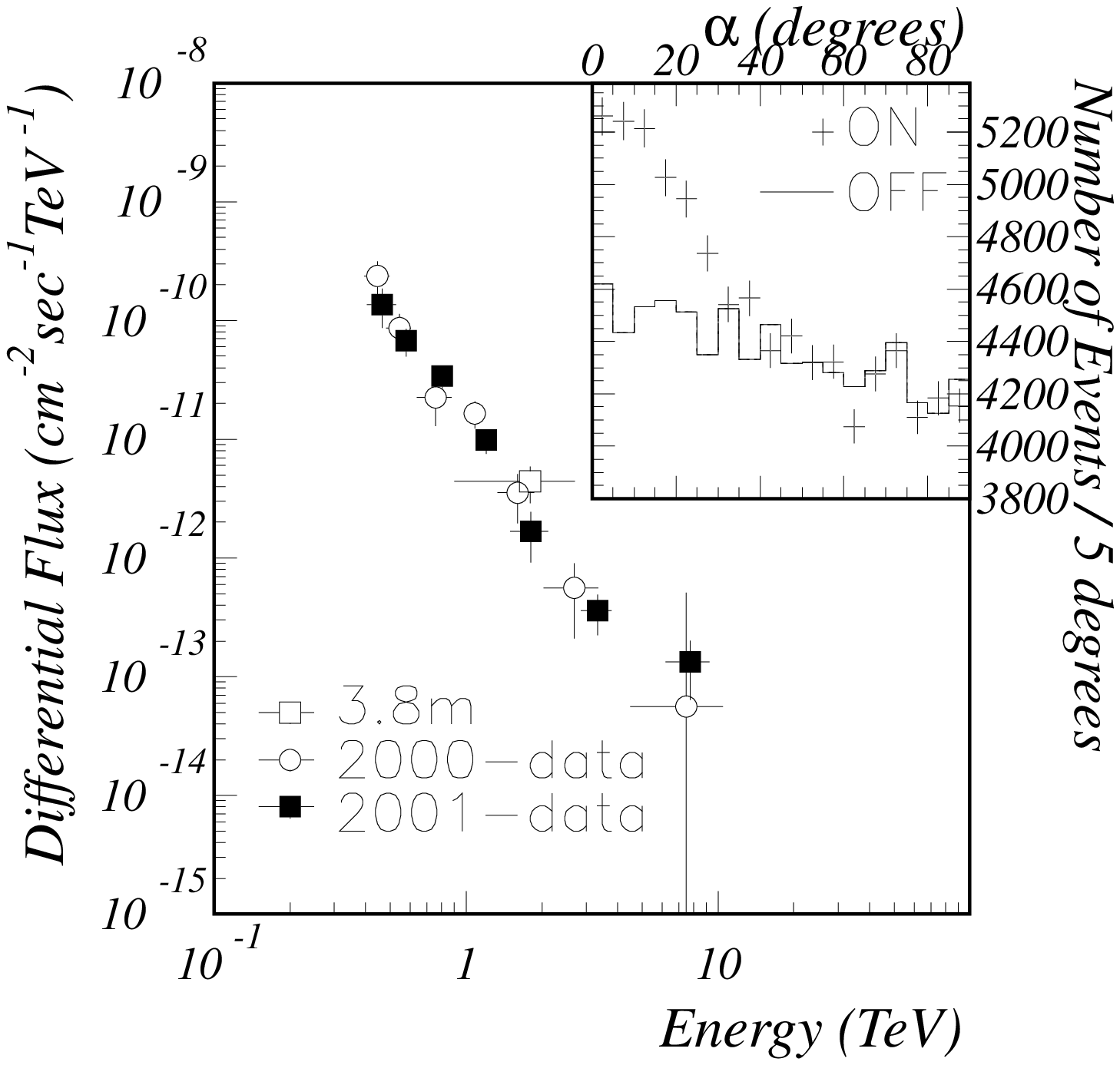}}
\caption 
{{\bf a:} Differential fluxes obtained by this experiment 
together with that of CANGAROO-I. 
Inserted graph is the excess events 
determined from the plots of image orientation angle.
}
\end{minipage}\hfill
\hspace{5mm}
\begin{minipage}[t]{6cm}
\resizebox{6cm}{!}{\includegraphics{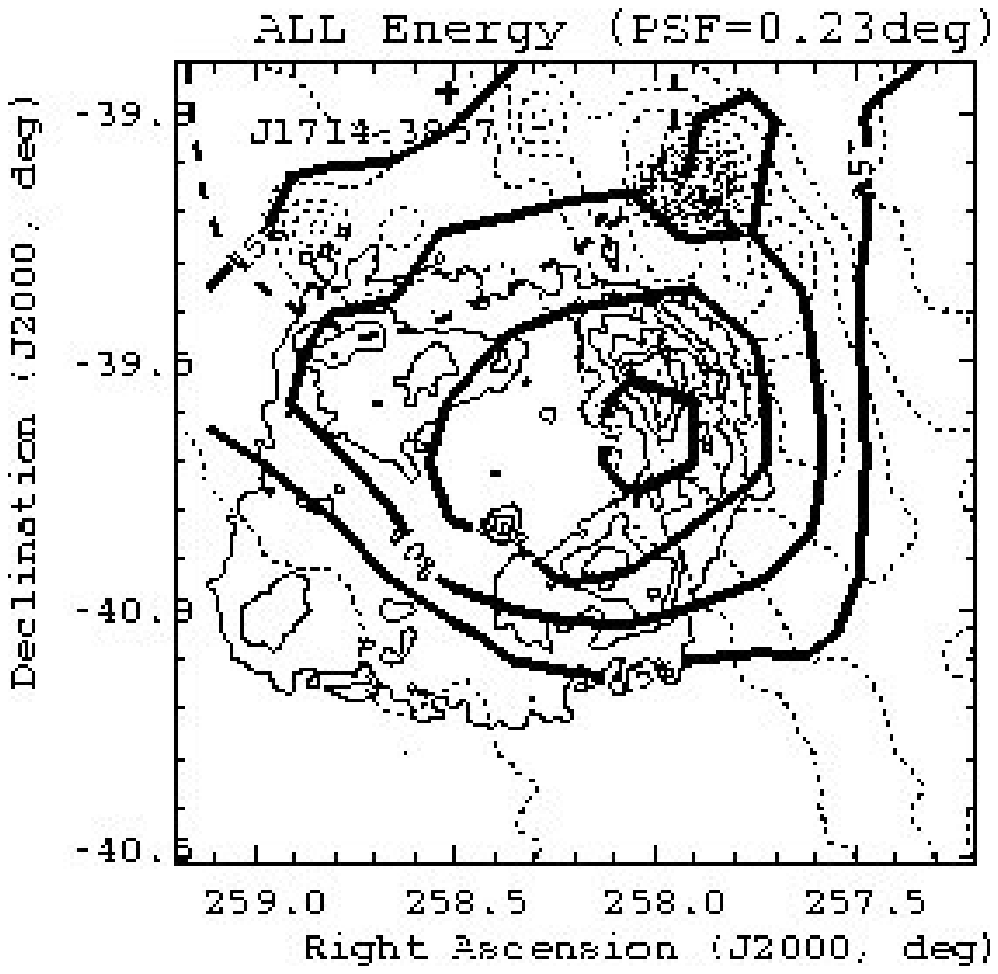}}
\caption 
{{\bf b:}  Profile of emission of TeV gamma rays around the NW-rim of RX J1713.7-3946
(solid thick contours).
}
\label{1713-dif}
\end{minipage}
\end{figure*}

Morphology of the gamma-ray emitting region is shown by the thick-
solid contours in Fig.{\ref{1713-dif}b,
 together with the synchrotron X-ray  ($\ge$ 2keV) 
contours by ASCA\cite{n19} and infrared ones from IRAS 100$\mu$m  
results\cite{n20} which possibly indicates the density distribution 
of the inter stellar matter. 
The observed TeV gamma-ray intensity peak coincides with the maximum point NW-
rim observed in X-ray, but 
the TeV gamma ray emission extends over the ASCA contours. 
A possible extension towards the CO cloud in the north-east\cite{n21} 
can also be seen.

The broad band energy spectrum is plotted in Fig.\ref{1713-allow}a
with theoretical predictions (described below).
Also in this figure, other data  has been shown  using data 
from the ATCA (Australia Telescope Compact Array)\cite{n22}, 
ASCA\cite{ref:35,n19},and EGRET\cite{n24}. 

\begin{figure*}
\begin{minipage}[t]{8cm}
\resizebox{8cm}{!}{\includegraphics{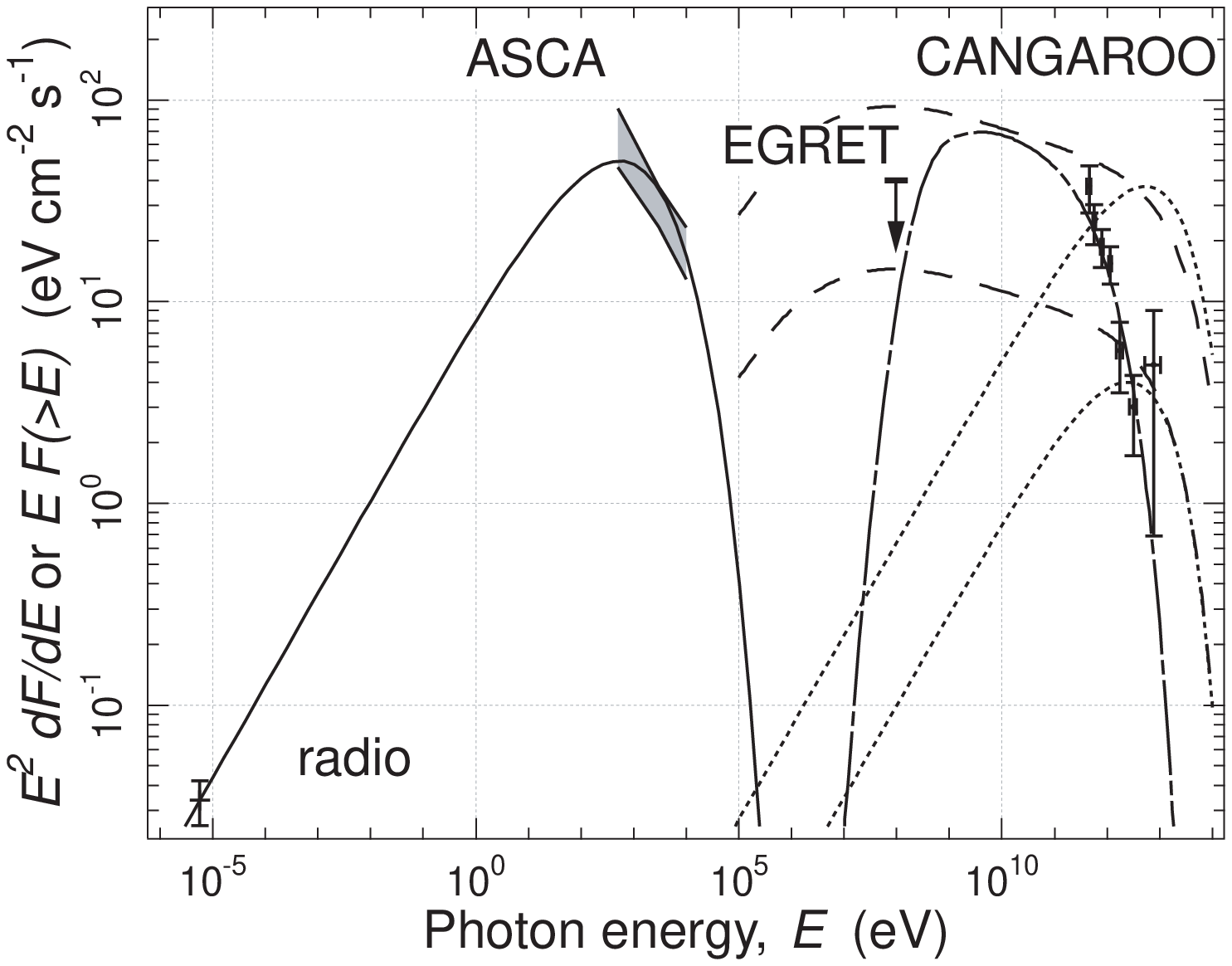}}
\caption
{{\bf a:} Multi-band emission  with models. 
TeV gamma-ray points are from this work. Lines 
show model calculations: synchrotron emission (solid line), Inverse Compton 
emission (dotted lines), bremsstrahlung (dashed lines) and emission from 
$\pi^{\circ}$decay (dot-dashed line).  
}
\end{minipage}\hfill
\hspace{5mm}
\begin{minipage}[t]{6cm}
\resizebox{6cm}{!}{\includegraphics{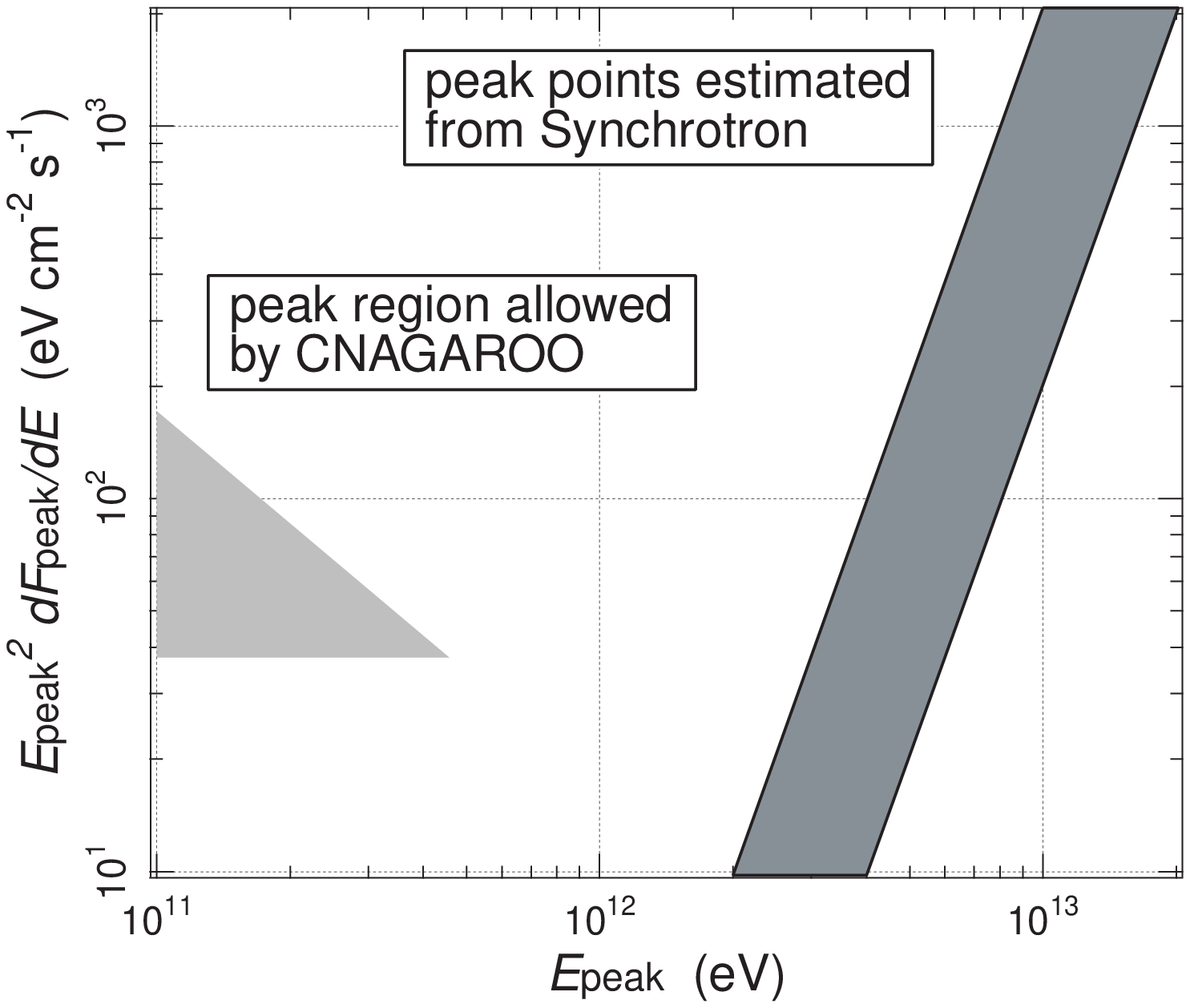}}
\caption
{{\bf b:} Allowed regions in parameter space (peak flux versus peak 
energy). The peak energies and fluxes for IC processes are allowed in the 
above parameter space, and also the allowed region from 
experimental flux 
corresponds to the lower-right corner of the shaded area. }
\label{1713-allow}
\end{minipage}
\end{figure*}

In order to explain the broad-band spectrum, 
three mechanisms,
the synchrotron/IC process, bremsstrahlung, and $\pi ^{\circ}$ decay 
produced by proton-nucleon collisions are considered, where 
the momentum spectra of incident particles (electrons and 
protons) are assumed to be

\begin{equation}
\frac{dN}{dp}
= N_{0} {\left(\frac{p}{mc} \right)}^{-\alpha}
\exp \left(-\frac{p}{pmax} \right)  [cm^{-2}eV^{-1}c]\ \ \ .
\end{equation}

\noindent based on the shock acceleration model.
The effects of acceleration limits from the age and size of the SNR and
energy losses of particles are included in the exponential term. 
Best-fit values of free parameters in (2) for the radio and X-ray 
fluxes due to synchrotron radiation from electrons  are obtained 
2.08\cite{n26} for $\alpha$, 126 for $ (p_{max}/TeV)(B/\mu G)^{0.5}$, 
where $c$ is the speed of light, 
and 2.00 for $ N_{\circ}(V/4\pi d^{2})(B/\mu G)^{(\alpha +1)/2} $, 
where $V$ is the volume of the radiation region and d is 
the distance from the Earth. 

Thus the resultant best fit is plotted in Fig.\ref{1713-allow}a 
(the solid line). 
We initially assumed the 2.7 K CMB as the seed 
photons for IC scattering. 
Calculated inverse Compton Spectra are  plotted 
with dotted lines in Figure 2 for two typical
magnetic field strengths, 3 and 20$\mu$G.
Note  that these models are far from  consistent with the 
observed sub-TeV spectrum. 
In general synchrotron/IC process gives a clear  correlation between the 
peak fluxes and its energies of synchrotron and IC emissions as a function 
of the strength of the magnetic field.
We investigated the allowed region of peak flux of IC emission 
taking into account the uncertainties of IR emission for the IC seed photons. 
The calculation was carried out both with and without this IR maximum 
energy emission in the IC process, 
assuming various incident-electron spectra. 
The hatched area in Fig.\ref{1713-allow}b is the resulting 
theoretically allowed region of the peak flux, and also 
the experimentally allowed region are plotted  by the shaded area. 
The predictions of synchrotron/IC models are obviously inconsistent with our 
experimental data by an order of magnitude. 
The bremsstrahlung spectrum was calculated assuming that it occurs in the 
same region as the synchrotron radiation. 
A material density of $\sim 300 protons/cm^{3}$ was assumed. 
The dashed lines in Fig.\ref{1713-allow}a, 
for magnetic fields of 3 and 20$\mu$G, are both 
inconsistent with our observation.

Thus, electron-based models fail to explain the observational results and so 
we examined $\pi^{\circ}$ decay models. 
The $\pi^{\circ}$s are produced in collisions of accelerated protons 
with interstellar matter. 
A model\cite{n25} adopting $\Delta$-resonance and scaling was used. 
We adopted parameters for Equation (2) of $\alpha$ = 2.08 and $p_{max}$ = 10 TeV, 
considering the plausible parameter regions of 
typical shock acceleration theory. 
The result is shown by the dot-dashed curve in Fig.\ref{1713-allow}a. 
The best fit parameters for the total energy of accelerated 
protons $E_0$ and matter density $n_0$ must satisfy 
$(E_0/10^{50} [ergs])\cdot (n_0 [protons/cm^{3}])\cdot (d/6 [kpc])^{-2} = 300 $, 
where $d$ is the distance to RX J1713.7-3946. 
A value of $E_0 \sim 10^{50} [ergs]$ gives $n_0$ 
of the order of 10 or $ 100 [protons/cm^{3}]$ 
for distances of 1 or 6 kpc, respectively. 
Both cases are consistent with the molecular column density estimated 
from Fig. 7 of ref.6. 
Thus the $\pi^{\circ}$ decay model alone readily 
explains our results, which 
provide the first observational evidence that protons are 
accelerated in SNR to at least TeV energies.

We can guess more  about the galactic cosmic-ray origin.
Using observed TeV gamma-ray flux, 
total energy of protons in the SNR was 
estimated to be $E_0 \sim 10^{50} erg$, 
assuming the distance of 6 kpc and target molecular cloud density of 
several times of $10^{2} protons/cm^{3}$. 
The energy input rate for proton acceleration estimated from this observation 
can represent the energetics of cosmic rays in the Galaxy. 
In other words, if such SNR are born every 100 years and accelerate protons, 
the SNR can provide all the energy of cosmic rays inside the Galaxy, 
where we adopt 1 $\times 10^{40} erg/sec $ 
for the luminosity of galactic cosmic-rays. 

\section{Summary}
Recent detections of TeV gamma rays from SNRs will obviously advance
the studies of the galactic cosmic-ray origin and the shock acceleration 
mechanism by  stimulating  multi-wave length observations for SNRs.
In particular combination of the morphological studies in both 
X-ray and gamma-rays will be a key for these  studies;
advanced X-ray telescope satellites, Chandra and Newton, are now providing
excellent images and spectroscopy, also the stereo observation
by the 10m-class IACTs will soon provide good quality
TeV gamma-ray images with the angular resolution of $\le$ 0.1 degree.
In southern hemisphere,  CANGAROO and H.E.S.S groups will begin the
stereo observations using 10m-class IACTs in 2002.

\end{document}